
\input amstex.tex
\documentstyle{amsppt}
\loadbold
\magnification=\magstep2
\pagewidth{11.3cm}
\pageheight{16cm}
\headline={\hfill\number\pageno}

\topmatter

\vskip 1cm

\title On Quaternions and Monopoles 
\endtitle
\author G.G. Emch  ${}^{\star}$ 
and A.Z. Jadczyk ${}^{\star\,\dagger}$
\endauthor
\affil ${}^{\star}$ Dept. of Mathematics, Univ. of Florida, 
Gainesville, FL 32611\\
${}^{\dagger}$ Institute of Theoretical Physics, University of 
Wroclaw
\endaffil

\footnotetext{\it{ 1992 Physics and Astronmy classification: 
03.65.Fd , 03.65.Bz , 03.65.Ca}}

\nologo
\abstract{It is shown that the quaternionic Hilbert space formulation 
of
quantum mechanics allows a quantization, based on a generalized
system of imprimitivity, that leads to a description of the motion of 
a 
quantum particle in the field of a magnetic monopole. The 
corresponding
Hamilton operator is linked to the theory of projective 
representations
in the weakened form proposed by Adler. }
\endabstract
\endtopmatter
\document
\flushpar {I.  INTRODUCTION}
\bigskip

Symmetries are one of the most powerful tools in theoretical physics. 
And yet there are few, if any, exact symmetries in Nature. Thirty 
years ago 
Hans Ekstein$^{1}$ 
addressed this problem by introducing the concept of 
\lq\lq presymmetry\rq\rq -- 
a pre--dynamical symmetry, that is being broken by dynamics and yet 
is evidenced in the algebra structure.  
Adler$^{2}$ 
introduced the concept of a
\lq\lq weak projective representation \rq\rq 
(WPR) and analyzed it within the framework of quaternionic quantum 
mechanics (see also $^{3,4}$  for the epistemological controversy which 
arose around this concept). In a recent note 
Adler and Emch$^{5}$
revisited the basic concepts of strong and weak projective 
representations from the point of view of Wigner's theorem 
$^{6}$ 
and the axiomatic formulation of quaternionic quantum mechanics 
extensively analyzed by one of us (GGE) more than thirty years ago
$^{7}$. 
Almost concommitantly, more than twenty years ago, following the 
original ideas of 
Ekstein$^{1}$, 
one of us (AZJ) introduced the concept of a generalized imprimitivity 
system (GIS) -- a concept which involves operator--valued multiplier 
as in WPR. 
In $^{8}$ 
the Stone -- von Neumann theorem was enhanced so as to also apply to
GIS's, and in 
$^{9}$, 
as an illustration, a GIS corresponding to a charged quantum particle 
in the field of Dirac's magnetic monopole was explicitly constructed. 
In the present paper these approaches are brought together, and we 
show that WPR's arise naturally from GIS's and that they correspond 
to symmetries that are only partially broken, with a remaining 
presymmetry (in the sense of Ekstein) holding only for an Abelian 
subalgebra of the algebra of all observables. We illustrate these 
concepts on the example of magnetic monopole quantum mechanics; we 
show that the, heretofore, somewhat mysterious half-spin properties
$^{10}\, $
-- by the very nature of the Clifford algebra 
$^{11} $
of ${(\Bbb E^3;-1,-1,-1)}$ -- 
naturally involve a quaternionic Hilbert space formulation.   
\bigskip
\bigskip

\flushpar {II.  MOTIVATING MODEL}
\bigskip 

Our model describes quantum kinematics and dynamics of a charged 
particle in the field of a magnetic monopole. The model is realized 
in a space of square integrable sections of a Hermitian quaternionic 
line bundle over 
${\Bbb R^3}\setminus \{0\}.$
The basic properties and notations relative to the field of 
quaternions 
$ {\Bbb H} $ 
and the quaternionic Hilbert space
$ {\Cal H}_{\Bbb H} = {\Cal L}^2 ( {\Bbb R}^3 , d^3 x ; {\Bbb H} )$ 
are reviewed in the Appendix. From a measure--theoretical point of 
view the Hilbert spaces 
${\Cal L}^2 ( {\Bbb R}^3 , d^3 x ; {\Bbb H} )$
and 
${\Cal L}^2 ( {\Bbb R}^3\setminus \{0\} , d^3 x ; {\Bbb H} )$
are naturally isomorphic and we will not make any distinction between
them until section IV, where the differential geometric aspects
of the construction will be discussed. 

$\bullet $ The position operators are defined, as usual, by 
$ [X_i \,\psi] (x) = x_i \,\psi (x) \, ;$ 
we denote by 
$\{ E(\Delta)|\Delta\subset{\Bbb R^3}\}$ 
their spectral family (see Appendix). Our model is spherically 
symmetric, with the rotation generators 
$M_i$ given by
$M_i=\epsilon_{ijk}x_j\partial_k-{1\over2}{\hat e}_i \, ,$
where
$ \epsilon_{ijk} $
is the totally antisymmetric tensor with 
$ \epsilon_{ijk} = 1 $ for $ ijk $ any cyclic permutation of the 
indices 
$ 123 \,  $ 
-- so that, e.g. 
$ \epsilon_{ijk} a_j \, b_k = (\bold a \times \bold b)_i \, $ -- 
and where 
$ e_1,e_2,e_3 $ are the three standard quaternion imaginary units.

For every $0\neq {\bold x}\in {\Bbb R}^3 \, $ let
$j(\bold x)$ the imaginary unit quaternion
$$j(\bold x)=\frac{\bold e\cdot \bold x}{\Vert \bold x\Vert} \quad . 
\tag 2.1a$$
$\bullet $  The linear operator $J={\hat j},$ i.e.:
$$ ( J \psi ) (\bold x) = j(\bold x) \psi (\bold x) 
\tag 2.1b $$
satisfies the two relations
$ J^* J = I = J J^* $ and $ J^* = - J \, , \, $ 
i.e. is  unitary and anti--hermitian; clearly, we also have 
$ J^2 = - I \, . $ 
Moreover $J$ is invariant under rotations and commutes with the 
position operators.

For every direction
$ \bold u \in S^2 = \{ \bold u \in {\Bbb R}^3 \mid \Vert \bold u 
\Vert = 1 \} \, , $
we construct an anti-hermitian operator
$ \nabla_{\bold u} $ given by the formula:
$$ \nabla_{\bold u} = \bold u \cdot \boldsymbol\partial + \frac12 
\frac{\bold e\cdot [\bold u\times \bold x]}{\Vert \bold x \Vert^2} \quad .
\tag 2.2 $$

$\bullet $  $ \nabla_{\bold u} $
generates a one-parameter unitary group
$ \{ U_{\bold u} (s) \mid s \in {\Bbb R} \} $ 
which satisfies, for all 
$ s \in {\Bbb R} $
and all Borel subsets 
$ \Delta \subset {\Bbb R}^3 \, : $
$$ U_{\bold u} (s) \, E (\Delta) \, U_{\bold u} (-s) = E(\Delta - s 
{\bold u} ) \, ;
\tag 2.3$$
or, infinitesimally:
$$[\nabla_i,x_j]=\delta_{ij} .$$ 
Thus 
$\nabla_{\bold u}$ 
generates translations in the direction 
$\bold u$ 
of the position variables. Moreover, we have 
$[M_i,\bold\nabla_j]=-\epsilon_{ijk}\nabla_k \, ,$ 
so that 
$\boldsymbol\nabla$ 
transforms as a vector under rotations.

$\bullet $ The unitary evolution defined by
$$U(t)=\exp (-JHt)\, \text{where}\, H = -\frac{1}{2m} \, 
\boldsymbol\nabla^2  \quad \text{and} \quad 
\boldsymbol\nabla^2 = \sum_{i=1}^3 (\nabla_i)^2  
\tag 2.4 $$
gives the evolution equations for the position operator
$ \bold X \, , $ namely : 	
$$ \overset \,{.}\to {X_i}  = -\frac{J}{m} \, \nabla_i  
\tag 2.5a $$
and 
$$ \overset \,{..}\to {X_i} =    \frac{1}{2m} \, \epsilon_{ijk}
( \overset \,.\to X_j \,B_k + B_j \, \overset \,.\to X_k ) 
\tag 2.5b $$
with
$$   [B_i \psi](\bold x) = \frac{1}{2}\frac{x_i}{\Vert \bold x \Vert 
^3} \psi (\bold x) \quad ,
\tag 2.6 $$
which correspond to the motion of a charged particle in the field of
a magnetic monopole.

$\bullet $ The translation generators do not commute:
$$ [\, \nabla_i \, , \, \nabla_j \, ] = 
-\frac12 \epsilon_{ijk} \frac{x^k}{\Vert \bold x \Vert ^3} \, J 
\tag 2.7 $$
which implies that the unitary operators
$ \{ U(\bold a) \mid \bold a \in {\Bbb R}^3 \} $
defined by
$ U(s\,\bold u) = U_{\bold u}(s) $ 
for all 
$ s \in {\Bbb R } $ and $ \bold u \in S^2 \, , $ 
i.e. for all 
$ s\,\bold u \in {\Bbb R}^3 \, , $
are only a WPR of the translation group in the sense of Adler.

$\bullet $ The following \lq\lq splitting\rq\rq relations are 
satisfied:
$$ 0=[X_i,J]=[\nabla_{\bold u},J]=[H,J]\, . \tag 2.8 $$
\bigskip
\bigskip

\flushpar {III.  DETAILS OF THE CONSTRUCTION}
\bigskip

The canonical quantization is given by the system of imprimitivity 
where
$$ V(\bold a) E(\Delta) V(-\bold a) = E(\Delta - \bold a) \quad 
\text{with} \quad
   [V(\bold a) \psi] (\bold x) = \psi (\bold x - \bold a) \quad ,  
\tag 3.1 $$
where
$ \{ V(\bold a) \mid a \in {\Bbb R}^3 \}$ is a continuous unitary 
representation with generators
$\partial_i\, . $ 
These generators correspond to covariant derivatives of the flat 
connection. In presence of an external magnetic field: vector 
potential enters into the connection form; covariant derivatives 
cease to commute; parallel transport becomes path dependent; 
translational symmetry is partially broken; and an operator--valued 
multiplier corresponding to an integral curvature enters into the 
group composition formula. In the present paper we want to draw
attention to the clarifying role played by the quaternions; we skip 
therefore any further heuristic motivation of the construction.

We define, for every $\bold a \in {\Bbb R}^3$ and for all $\bold x\in 
{\Bbb R}^3$ not colinear with 
$\bold a$
$$w(\bold a;\bold x)=\frac{1}{\sqrt{2}}\left(\sqrt{1+
\frac{{\Vert \bold x \Vert}^2+\bold a\cdot \bold x}
{\Vert \bold x\Vert \Vert \bold x+\bold a\Vert}}+
j(\bold x\times \bold a)\sqrt{1-\frac{{\Vert \bold x \Vert}^2
-\bold a\cdot \bold x}{\Vert \bold x\Vert\, \Vert \bold x
+\bold a\Vert}}\right), 
\tag 3.2$$
and let $W(\bold a)$ denote the bounded linear operator ${\hat 
w}(\bold a;\cdot)$, that is
$$(W(\bold a)\psi)(\bold x)=w(\bold a;\bold x)\psi(\bold x)\quad 
\text{ a.e. }
\tag 3.3$$
(see Appendix).

It can be verified that:

$\bullet $ $w(\bold a;\bold x)w(\bold a;\bold x)^\star=1$ a.e., and 
thus $W(\bold a)$ are unitary operators.
They commute with the position observables.

$\bullet $ $w(\bold a;\bold x)$ satisfy the cocycle relations
$$w(t\bold a,\bold x+s\bold a)w(s\bold a,\bold x)w(s\bold a,\bold 
x)=w((s+t)\bold a,\bold x),\quad {\text {a.e.}}\, . $$ 
For every $\bold a\in {\Bbb R}^3 \, , $ 
define
$$U(\bold a)=V(\bold a)W(\bold a)$$
and for each $\bold u\in S^2$ and $s\in {\Bbb R}\, , $ 
let
$$U_{\bold u}(s)=U(s\bold u)\, ;$$

$\bullet $  $\{ U_{\bold u}(s)\vert s\in {\Bbb R}\}$ is a continuous 
unitary group representation of 
${\Bbb R}\,$ 
whereas 
$\{U(\bold a)\vert a\in {\Bbb R}^3\}$
will only be a weak projective representation -- see below. 

$\bullet $ By a direct computation one verifies that, for every 
direction $\bold u\, ,$ the infinitesimal generator 
$\nabla_{\bold u}$ of $U_{\bold u}(t)$ is given by $(2.2)\, . $

$\bullet $ $U(\bold a)$ satisfy the imprimitivity relations $(2.3)$; 
it follows that for all
$\bold a,\bold b\in {\Bbb R}^3 :$
$$U(\bold a)U(\bold b)=U(\bold a+\bold b)M(\bold a,\bold b) 
\tag 3.4 $$
with $M(\bold a,\bold b)$ commuting with $ E(\Delta) $
for all Borel subsets 
$ \Delta \subseteq {\Bbb R}^3 \, . $ 
Thus $M(\bold a,\bold b)$ 
are of the form 
$(M(\bold a,\bold b)\psi)(\bold x)=
m(\bold a,\bold b;\bold x)\psi(\bold x)$. 
In Adler's notation $^1\, , $
this reads 
$M(\bold a,\bold b)=\int |\bold x>m(\bold a,\bold b;
\bold x)<\bold x|\, d^3x\, .$
Upon writing 
$m(\bold a,\bold b;\bold x)$ 
in terms of $w(\bold a;\bold x)$ we find: 
$$m(\bold a,\bold b;\bold x)=w(\bold a+\bold b;\bold x)^\star w(\bold 
a;\bold x+\bold b)w(\bold b;\bold x)\in {\Bbb H} \, .
\tag 3.5 $$  
In fact, by a direct calculation, we receive:
$$m(\bold a,\bold b;\bold x)=\exp (J\Phi(\bold a,\bold b;\bold x))\, 
,\tag 3.6 $$ where $\Phi(\bold a,\bold b;\bold x)$ is the flux
of the monopole magnetic field through the flat triangular surface 
spanned by the vertices 
$(\bold x,\bold x+\bold a,\bold x+\bold a+\bold b)\, .$ 
The cocycle formula for $M(\bold a,\bold b)$ expressing
associativity of the operator product 
$(U(\bold a)U(\bold b))U(\bold c)=U(\bold a)(U(\bold b)U(\bold c))$ 
is then interpreted as stating that the flux through the closed 
tetrahedron spanned by the edges 
$(\bold x,\bold x+\bold a,\bold x+\bold a+\bold b,
\bold x+\bold a+\bold b+\bold c)$ 
is an integer multiple of 
$2\pi\, $
which is automatically satisfied by the magnetic field of the 
monopole - see (2.6).

\bigskip
\bigskip

\flushpar {IV.  DISCUSSION}
\bigskip

Our magnetic monopole model  is constructed in a quaternionic Hilbert 
space
${\Cal H}_{\Bbb H}$,
yet it admits a commuting antiunitary involution $J$ and thus reduces, 
de facto, to a complex Hilbert space model in $ {\Cal H}_{\omega }$. 
The phenomenon of a "weak projective representation", in the sense 
implied by Adler, here for the translation group, shows up in both 
the quaternionic space and in the complex reduction. This is
because the "twisted translations" $U(\bold a)$ commute with $J .$ A 
differential geometric interpretation of the construction is helpful 
in order to understand at a deeper level what is really going on here. 
The Hilbert space 
$ {\Cal H}_{{\Bbb H}} = {\Cal L}^2 ( {\Bbb R}^3, d^3 x ; {\Bbb H}) $
can be considered as a Hilbert space of square integrable sections of 
a trivial Hermitian complex line bundle $F$ over 
${\Bbb R}^3\backslash\{\bold 0\}.$ 
Removing the origin results in no measurable theoretic consequences;
this removal however does have differential geometric sequels. Our 
operators 
$\nabla_{\bold u}$
define a Hermitian connection in $F$. The curvature two--form 
$\Omega\, , $ with values in
the Lie algebra $su(2)$ is given by the formula 
$$\Omega^r = 
- \frac12 \epsilon_{ijk} \frac{x^k x^r}{\Vert \bold x \Vert ^4} \, 
dx^i\wedge dx^j \, . 
\tag 4.1$$

The fact that the operator $J$ defined by (2.1) commutes with 
$\nabla_{\bold u}$ 
can be interpreted as stating that the map 
$\bold x\rightarrow j(\bold x)$ 
is a parallel section of the bundle of quaternionic right-linear 
endomorphisms of $F.$ The formula (A.10) defining 
$ {\Cal H}_{\omega }$ 
describes, de facto, a construction of a Hermitian complex subbundle 
$F_{\omega}$ of $F$ 
which reduces the connection 
$\boldsymbol\nabla .$
The complex Hilbert space 
$ {\Cal H}_{\omega }$ 
consists of square--integrable sections of the bundle 
$F_{\omega}.$ Because $J$ is invariant under rotation, it follows 
that the rotation group acts covariantly on 
$F_{\omega}$ and unitarily on $ {\Cal H}_{\omega }$
and is a two-valued representation of 
$SO(3)$ 
corresponding to spin one--half. At first sight, it might appear 
somewhat surprising that we can have spin one--half in a Hilbert 
space of complex, one-component, functions. To answer this puzzle, 
we note that the bundle 
$F_{\omega}$ 
is non--trivial. It admits no continous, nowhere zero, sections -- 
it carries a spin one--half "kink". To see that the bundle is 
nontrivial we compute the simplest topological invariant, that is 
its first Chern class. In our case it is the integral of the curvature 
two--form 
$\kappa\, , $ with now:
$$ \kappa = 
-\frac12 \epsilon_{ijk} \frac{x^k}{\Vert \bold x \Vert ^3} 
dx^i\wedge dx^j \, ,
\tag 4.2 $$
over the sphere $S^2$ - the result is $2\pi$ which
proves that the bundle is non--trivial. 

While $\nabla_{\bold u}$ (and thus $H$) have a simple explicit form, 
as globally defined differential operators  on a dense domain of 
differentiable functions in
${\Cal H}_{{\Bbb H}}\, ,$ 
which is built out of sections of a trivial vector bundle over 
${\Bbb R^3}\setminus\{\bold 0\}\, $ -- 
their restriction to 
$ {\Cal H}_{\omega }$ 
cannot be so written; this is due to the fact that
$ {\Cal H}_{\omega }$ 
is defined in terms of sections of a non--trivial subbundle over  
${\Bbb R^3}\setminus\{\bold 0\}\, .$ 
If we were to force an explicit expression for the covariant 
derivative in the reduced bundle, a string--like singularity would 
have to appear -- a one--point singularity on each sphere of the 
constant radius 
$0<r\in {\Bbb R}\, .$ 
Hence the definite advantage of working with the quaternionic Hilbert 
space ${\Cal H}_{{\Bbb H}}$. 

Working with singularity--free formulation does not depend by itself 
on the quaternionic structure -- we could use as well a 
${\Bbb C}^2$ 
bundle; nevertheless, the full gauge freedom of the theory is 
manifest only from within a quaternionic perspective.   

Let us, finally, comment upon the relations between the present work 
and GIS'es studied in Refs.$^{8,9}\, .$ To define a GIS we need an 
action of a group $G$ on a space 
$X\, .$ 
In the most regular case, $G$ is a Lie group acting differentiably 
on a manifold $X.$  A GIS is then defined by the relations:
$$U(g)E(\Delta)U(g)^{\star}=E(\Delta g) 
\tag 4.3a$$
$$U(g)U(h)=U(gh)M(g,h) 
\tag 4.3b$$
$$M(g,h)=\int_X m(g,h;x) dE(x)\, , 
\tag 4.3c$$
where $g:\rightarrow U(g)$ is a continuous map from $G$ into unitary 
operators acting on the Hilbert space 
${\Cal H}\, ,$ and $m(g,h;x)$ 
commute with the spectral measure. In the example discussed in the 
present paper $X$ is the three--dimensional Euclidean space 
$\bold E^3\, ,$ $G$ 
is its translation group, and
$m(g,h;x)$ are quaternionic valued. It is seen that a GIS always 
gives rise to a WPR in the sense of Adler. It is however to be 
remarked that the very concepts of a GIS  (and also of WPR) has 
little to do with the field over which the Hilbert space is defined. 
The concept applies to real, complex or quaternionic Hilbert spaces 
as well. 

The idea of a "presymmetry" -- that is of a symmetry group which is 
partially broken by the dynamics, but yet still corresponds to a full 
symmetry group on an Abelian subalgebra -- is quite naturally 
supported by the GIS framework; in contrast, the a priori 
mathematically more general concept of WPR leaves open the choice of 
the sub--algebra necessary to the physical interpretation of the group 
of (pre--)symmetries; the formulation in terms of GIS seems therefore 
to help specify physically the choice latent in the WPR formulation.

\bigpagebreak
\flushpar {\bf{Acknowledgements.}}
The authors thank Dr. S.L. Adler for rekindling their interest
in quaternionic quantum mechanics, and thus in the problem discussed 
in this paper. One of us (A.J) thanks the Kosciuszko Foundation for 
financial support that enabled our collaboration. He also thanks 
Prof. John Klauder and the Mathematics Department of the University 
of Florida for their hospitality.

\bigskip
\bigskip

\flushpar {APPENDIX}
\bigskip

The field
$ {\Bbb H} $ of the (real) quaternions is obtained upon equiping 
the 4--dimensional real vector space
$$ {\Bbb H} = \{\, q = \sum_{\mu = 0}^3 \, q^{\mu} e_{\mu} \mid 
   a^{\mu} \in {\Bbb R} \,\}
\tag A.1 $$
with the non--commutative multiplication it inherits from
$$ e_o \, q = q\, e_o \,\,\forall\,\, q \in {\Bbb H} \, ; \,
e_i \, e_j = -\delta_{ij}e_0+\epsilon_{ijk}\, e_k\, , \quad i,j,k = 
1,2,3 . 
\tag A.2 $$
$ {\Bbb H } $ is equiped
with the involution
$$ q = \sum_{\mu}^3 \, a^{\mu} e_{\mu} \to 
 q^* = \sum_{\mu}^3 \, a^{\mu} e_{\mu}^* \quad \text{with} \quad
e_o^* = e_o \quad \text{and} \quad e_i^* = - e_i \quad .
\tag A.3 $$ 
Note that 
$ SU(2,{\Bbb C}) $ 
is isomorphic to 
$\{\, q \in {\Bbb H} \mid  q^* \, q = e_0 \, \} \, , $
with the isomorphism given by the identification
$$ e_o = \pmatrix 1 & 0 \\
                 0 & 1
         \endpmatrix \quad
   e_1 = \pmatrix 0 & -i \\
                 -i & 0
         \endpmatrix \quad
   e_2 = \pmatrix 0 & -1 \\
                  1 & 0
        \endpmatrix \quad
   e_3 = \pmatrix -i & 0 \\
                  0 & i
        \endpmatrix 
$$
i.e.
$$ e_o = I \quad \text{and} \quad e_k = - i \sigma_k 
\tag A.4 $$
where the 
$ \sigma_k $ are the three Pauli matrices. 
For any such quaternion, the map
$$ \alpha_{\omega} : q \in {\Bbb H} \mapsto \omega^*\, q\, \omega \in 
{\Bbb H}
\tag A.5 $$
is an automorphism of 
$ {\Bbb H} \, ,$
and every automorphism of
$ {\Bbb H} $
can in fact be implemented in this manner. In particular, if
$ \omega $
is an imaginary unit, i.e.
$$ \omega^* = - \omega\qquad
\text{and}\qquad
\omega^* \omega = e_o \, , \tag A.6$$
then
$$ \alpha_{\omega} [q] = q \quad \text{iff} \quad
   q \in {\Bbb C}_{\omega} = 
   \{\, u \, e_o + v \, \omega \mid u,v \in {\Bbb R} \,\} \quad .
\tag A.7 $$ 
Note that
$ {\Bbb C}_{\omega} $
inherits from
$ {\Bbb H} \, , $
the structure of the field
$ {\Bbb C} $
of the complex numbers.

We will henceforth use the notations
$ 1 =  e_o \, ,$
and
$ e = (e_1, e_2, e_3)  \, ,$
and for 
$ x \in {\Bbb R}^3 \, , $
$ x \cdot e =  \sum_{i=1}^3 x^i e_i \, . $
Note that  
$ q^*q = \Vert q \Vert ^2\, $ defines the quaternion norm, and 
that
$ ( x \cdot e )^* ( x \cdot e ) = \Vert x \Vert ^2 =  
\sum_{i=1}^3 (x^i)^2 \, .$

\vskip .3cm

The Hilbert space 
$ {\Cal H}_{{\Bbb H}} = {\Cal L}^2 ( {\Bbb R}^3, d^3 x ; {\Bbb H}) 
$
is the space of ``functions'' 
$ \psi : {\Bbb R}^3 \mapsto {\Bbb H} \, , $
square--integrable with respect to Lebesgue measure
$ d^3 x \, . $ Its vector space structure is defined with 
multiplication by 
scalars written from the right:
$$ [\psi q ] (x) = \psi (x) \, q \, , 
\tag A.8 $$
and the scalar product is given by:
$$ (\varphi , \psi ) = 
   \int_{{\Bbb R}^3} \, d^3 x \, \varphi (x)^* \, \psi (x) \quad .
\tag A.9 $$
It is linear in its {\it second} factor, and skew adjoint; hence
$ (\varphi q_1 , \psi q_2 ) = q_1^* \,(\varphi , \psi)\, q_2 \, . 
$

The linear operators
$ A : {\Cal H}_{{\Bbb H}} \to {\Cal H}_{{\Bbb H}} $
are denoted with left action, so that
$ A (\psi q ) = ( A \psi ) q = A \psi q \, . $ 
The adjoint is defined, as usual, by
$ ( \varphi, A^* \psi) = ( A \varphi , \psi ) $
$ \forall \,\, \varphi , \psi \in {\Cal H}_{{\Bbb H}} \, . $

Let $E$ be the spectral family 
$ [E (\Delta) \psi ] (x) = \psi (x) \chi_{\Delta} (x) $
where 
$ \Delta $
runs over all Borel subsets of
$ {\Bbb R}^3 \, ,  $
and 
$ \chi_{\Delta} $ 
is the indicator function of 
$ \Delta \, .$ 

We denote by ${\hat e}_i$ the linear anti-hermitian on ${\Cal 
H}_{{\Bbb H}}\, $
defined by left quaternion multiplication
$({\hat e}_i)\psi (x)=e_i \psi(x)\, .$ 

More generally, for each bounded measurable function $f:{\Bbb 
R}^3\rightarrow {\Bbb H} $
let $\hat f$ denote the bounded linear operator on 
$ {\Cal H}_{{\Bbb H}}$ defined by
$$({\hat f}\psi)(x)=f(x)\psi(x)\quad \text{a.e.}$$

The (real) commutant of $E$ consists then exactly of the operators of 
the form
$\hat f$.

For any unitary and anti--hermitian operator
$ J $ 
and any fixed imaginary unit
$ \omega \, , $
let
$$ {\Cal H}_{\omega } = \{\,\psi \in {\Cal H}_{{\Bbb H}} \mid J\,\psi 
= 
    \psi \, \omega\ \} \quad .
\tag A.10 $$
Note that
$ {\Cal H}_{\omega } $
inherits from
$ {\Cal H}_{{\Bbb H}} $ 
the structure of a complex Hilbert space over
the copy $ {\Bbb C}_{\omega } \, $ -- see $(A.7)$ -- of the field of 
complex numbers.
Specifically,
$ \varphi , \psi \in  {\Cal H}_{\omega } $
and
$ z \in {\Bbb C}_{\omega } $  
imply
$ \varphi + \psi \in  {\Cal H}_{\omega } \, ,$   
$ (\varphi , \psi ) \in  {\Bbb C}_{\omega } \, ,$   
and
$ \psi z \in  {\Cal H}_{\omega} \, . $
Furthermore, for every
$ \psi \in  {\Cal H}_{{\Bbb H}} \, , $
and every imaginary unit
$ \tilde\omega  $
such that
$ \tilde\omega \, \omega = - \omega \, \tilde\omega \, , $
there exists a unique pair
$$ \psi_1 , \psi_2 \in {\Cal H}_{\omega } \quad \text{such that} 
\quad
\psi = \psi_1 + \psi_2 \, \tilde\omega \quad ;
\tag A.11 $$
specifically
$$ \psi_1  =  \frac{1}{2}  ( \psi - J\,\psi\,\omega ) \quad 
\text{and} \quad
   \psi_2  = - \frac{1}{2} ( \psi + J\,\psi\,\omega )\,\tilde\omega 
\quad .
\tag A.12 $$
Note that as vectors in
$ {\Cal H}_{{\Bbb H}} \, , $ 
$ \psi_1 $ and $ \psi_2 $ 
are mutually orthogonal. Therefore the ``splitting'' 
(or ``dimension--doubling'')
$$ \psi \in {\Cal H} _{{\Bbb H}}\mapsto \pmatrix \psi_1 \\
                                      \psi_2
                             \endpmatrix         \in 
    {\Cal H}_{\omega} \oplus {\Cal H}_{\omega}   
\tag A.13 $$
is a bijective isometry.



\Refs

\ref\no 1
\by H. Ekstein
\paper
\jour Phys. Rev.
\vol 153
\yr 1967
\pages 153
\moreref
\jour Phys. Rev.
\vol 184
\pages 1315
\endref


\ref\no 2
\by S.L. Adler
\book Quaternionic Quantum Mechanics and Quantum Fields
\publ Oxford U.P.
\publaddr New York
\yr 1995 
\moreref
\by S.L. Adler
\paper Projective Group Representations
\jour J. Math. Phys.
\vol 37
\yr 1996
\pages 2352--2360
\endref

\ref\no 3
\by G.G. Emch
\paper Comments on a Recent Paper by S. Adler ...
\jour J. Math. Phys.
\vol 37
\yr 1996
\pages 6582--6585
\endref

\ref\no 4
\by S.L. Adler
\paper Response to the Comments by G. Emch ...
\jour J. Math. Phys.
\vol 37
\yr 1996
\pages 6586--6589
\endref

\ref\no 5
\by S.L. Adler and G.G. Emch
\paper A Rejoinder on quaternionic projective representations
\jour J. Math. Phys.
\vol 38
\yr 1997
\pages 4758--4762
\endref


\ref\no 6
\by G. Emch and C. Piron
\paper Note sur les sym\'etries en th\'eorie quantique
\jour Helv. Phys. Acta
\vol 35
\yr 1962
\pages 542-543
\moreref
\paper Symmetry in Quantum Theory
\jour J. Math. Phys.
\vol 4
\yr 1963
\pages 469--473
\endref

\ref\no 7
\by G.G. Emch
\paper M\'ecanique quantique quaternionienne et Relativit\'e 
restreinte. I.
\jour Helv. Phys. Acta
\vol 36
\yr 1963
\pages 739--769
\moreref
\paper II. 
\jour Helv. Phys. Acta
\vol 36
\yr 1963
\pages 770--788
\moreref
\paper Representations of the Lorentz Group in Quaternionic Quantum 
Mechanics
(presented at the Lorentz Group Symposium, Summer 1964
\inbook Lectures in Theoretical Physics, Vol. VIIa  
\publ W.E. Brittin, ed., U.Colorado Press,
\publaddr Boulder CO
\yr 1964
\pages 1--36 
\endref

\ref\no 8
\by A. Jadczyk
\paper On a theorem of Mackey, Stone and von Neumann for projective 
imprimitivity systems  
\jour Techn. Rep. IFT, Uni. Wroclaw, no. 
\vol 328 
\yr 1975 
\endref

\ref\no 9
\by A. Jadczyk
\paper Magnetic charge quantization and generalized imprimitivity 
systems
\jour Int. J. Theor. Phys.
\vol 14
\yr 1975
\pages 183--192
\endref

\ref\no 10
\by A.S. Goldhaber
\jour Phys.Rev.
\vol 140
\yr 1965
\pages 1407--1413
\endref

\ref\no 11
\by I.R. Porteous
\book Clifford Algebras and the Classical Groups
\publ Cambridge Univ. Press 
\publaddr Cambridge
\yr 1995
\endref


\endRefs
\enddocument

\end